\documentclass[aps,prb,twocolumn,superscriptaddress,showpacs,ams,floatfix]{revtex4}
\usepackage{color}
\usepackage{graphicx}
\usepackage{bm}
\usepackage{amssymb} 
\pdfoutput=1

\newcommand{\beq}   {\begin{equation}}
\newcommand{\eeq}   {\end{equation}}
\newcommand{\ba}   {\begin{eqnarray}}
\newcommand{\ea}   {\end{eqnarray}}

\newcommand{\leads}   {\mbox{\scriptsize leads}}
\newcommand{\coupling}   {\mbox{\scriptsize coupling}}
\newcommand{\MottIns} {\mbox{\scriptsize MI}}

\newcommand{\Equilib} {\mbox{\scriptsize eq}}
\newcommand{\BandIns} {\mbox{\scriptsize BI}}

\begin{document}

\title{Real-time dynamics of particle-hole excitations in Mott insulator-metal junctions.}
\author{Luis G.~G.~V. Dias da Silva}
\affiliation{Department of Physics and Astronomy, University of Tennessee, Knoxville, Tennessee 37996}
\affiliation{Materials Science and Technology Division,
Oak Ridge National Laboratory, Oak Ridge, Tennessee 37831}
\author{Khaled A. Al-Hassanieh}
\affiliation{Theoretical Division, Los Alamos National Laboratory, Los Alamos, New Mexico 87545}
\author{Adrian E. Feiguin}
\affiliation{Department of Physics and Astronomy, University of Wyoming, Laramie, Wyoming 82071,
USA}
\author{Fernando A. Reboredo}
\affiliation{Materials Science and Technology Division,
Oak Ridge National Laboratory, Oak Ridge, Tennessee 37831}
\author{Elbio Dagotto}
\affiliation{Department of Physics and Astronomy, University of Tennessee, Knoxville, Tennessee 37996}
\affiliation{Materials Science and Technology Division,
Oak Ridge National Laboratory, Oak Ridge, Tennessee 37831}

\date{\today}

\begin{abstract}
Charge excitations in Mott insulators (MIs) are distinct from their
band-insulator counterparts and they can provide a mechanism for energy
harvesting in solar cells based on strongly correlated electronic materials. In
this paper, we study the real-time dynamics of holon-doublon pairs in a MI
connected to metallic leads using the time-dependent density matrix
renormalization group method. The transfer of charge across the MI-metal
interface is controlled by both the electron-electron interaction strength
within the MI and the voltage difference between the leads. We find an overall
enhancement of the charge transfer as compared to the case of a
(noninteracting) band insulator-metal interface with a similar band gap.
Moreover, the propagation of holon-doublon excitations within the MI
dynamically changes the spin-spin correlations, introducing time-dependent
phase shifts in the spin structure factor.
\end{abstract}




\maketitle

\section{Introduction}
\label{sec: Intro}

The dynamics of excitations in strongly correlated electronic materials (SCEMs)
(Ref. \onlinecite{Tokura:462:2000}) has been the subject of intense study in
recent years. In a vast class of SCEMs, the ground state is a Mott insulator
(MI), characterized by strong on-site repulsive interactions and a charge gap
in the density of states. A paradigmatic model describing this behavior is the
one-dimensional (1D) Hubbard model at half-filling, in which elementary charge
excitations involve either doubly occupied (doublons) or empty (holons)
electronic sites. A doublon-holon pair is thus a charge neutral excitation,
which, depending on the relative strength of the on-site and nearest-site
repulsion, can either form a bound state (a ``Hubbard exciton") or remain
decoupled. These excitonic states in Hubbard-type systems have attracted
considerable attention involving both theoretical
\cite{Barford:205118:2002,Jeckelmann:75106:2003,Matsueda:153106:2005} and
experimental \cite{Kim:177404:2008,Gossling:75122:2008,Matiks::2009}
investigations.

A particular framework where the behavior of doubly occupied site excitations
has acquired considerable relevance is in experiments involving cold atomic
fermionic mixtures in optical lattices.\cite{Bloch:885:2008} While these
experiments are performed under controlled conditions and allow for a
substantial degree of tunability, the ability to properly probe and
characterize these systems is limited. A recently proposed technique to probe
the spectrum of a Mott insulating state consists of dynamically creating double
occupancies by modulating the lattice depth with a time-dependent optical
potential.\cite{Kollath:50402:2006} When the lattice depth exceeds the Mott
gap, double occupancy becomes favorable and, by optical methods, it is possible
to determine the appearance of the gapped mode.\cite{Joerdens:34:2008,
Huber:65301:2009} In addition, lattice-modulation spectroscopy has been
proposed as a method to detect antiferromagnetic ordering and probe the nature
(coherent or incoherent) of quasiparticle excitations.
\cite{Sensarma:35303:2009} The decay mechanisms associated to the doublon
excitations bring up fundamental questions about thermalization and
non-equilibrium dynamics.
\cite{Rosch:265301:2008,Heidrich-Meisner:41603:2009,Strohmaier::2009}

In addition, technological applications may potentially arise from exploiting
charge excitations in Mott insulators as a way to devise SCEM-based solar
cells, which can offer structural and optical advantages over current devices
made with semiconductor materials. One-dimensional Mott insulators (such as
$\mathrm{Sr_2CuO_3}$)  are particularly known for strong nonlinear optical
response effects\cite{Kishida:929:2000,Kishida:177401:2001} that can be
exploited in this context. The efficiency of SCEM-based  solar  cells will
depend on several factors, including the performance of Mott insulator-metal
junctions where the photocurrent will be  generated.  A  crucial question is
whether charge excitations in   the  MI will  be able  to properly transfer the
charge into the metallic   contacts, thus establishing a  steady-state
photocurrent.

Several questions arise in these regards: (i) Are the charge excitations
long-lived? In other words, what  are the effective decaying  channels for  the
holon-doublon excitations into  spin excitations inside the MI region? (ii)
What are the effects of the dynamics  of these excitations on the original
correlations in the MI region? (iii) Can these charge excitations propagate
across an     interface with a non-correlated material?

Point    (i)  was addressed     by some  of  us  in  Ref.
\onlinecite{Al-Hassanieh:166403:2008}, where the real-time dynamics of a
holon-doublon pair was studied in a 1D Mott insulator.  In that previous effort
it was found that  the decay to spin excitations in the underlying spin
background is inefficient, and the pair is long-lived. The weak decay to
spin-only excitations is particularly telling since the low-lying excitations
for the 1D Hubbard model at half-filling carry spin (spinon) and charge
(holons) quantum numbers separately and propagate with different characteristic
velocities, as described by the Tomonaga-Luttinger liquid picture. The ensuing
spin-charge separation of these low-energy modes has been long studied and the
real-time dynamics has been explored in recent
efforts.\cite{Kollath:176401:2005,Ulbricht:5:2009}

Points (ii) and (iii) will be addressed in this paper. Here, we study the
charge and spin dynamics  across a MI/metal   junction,  as doublon-holon
excitations are created within the MI. A possible mechanism for such
excitations arises from optical absorption at energies on the order of the Mott
gap, producing exciton-like states. We study the time evolution of a localized
doublon-holon pair initially created within the MI region. We consider a regime
with weak nearest-neighbor electron-electron repulsion relative to the kinetic
energy within the MI (or, equivalently, weak holon-doublon attraction), leading
to a spatial dissociation of the doublon-holon pair, with both excitations
eventually reaching the MI-metal boundaries.

\begin{figure}[tp]
\includegraphics*[height=0.5\columnwidth,width=1.0\columnwidth,bb=0 0 544 231,clip]{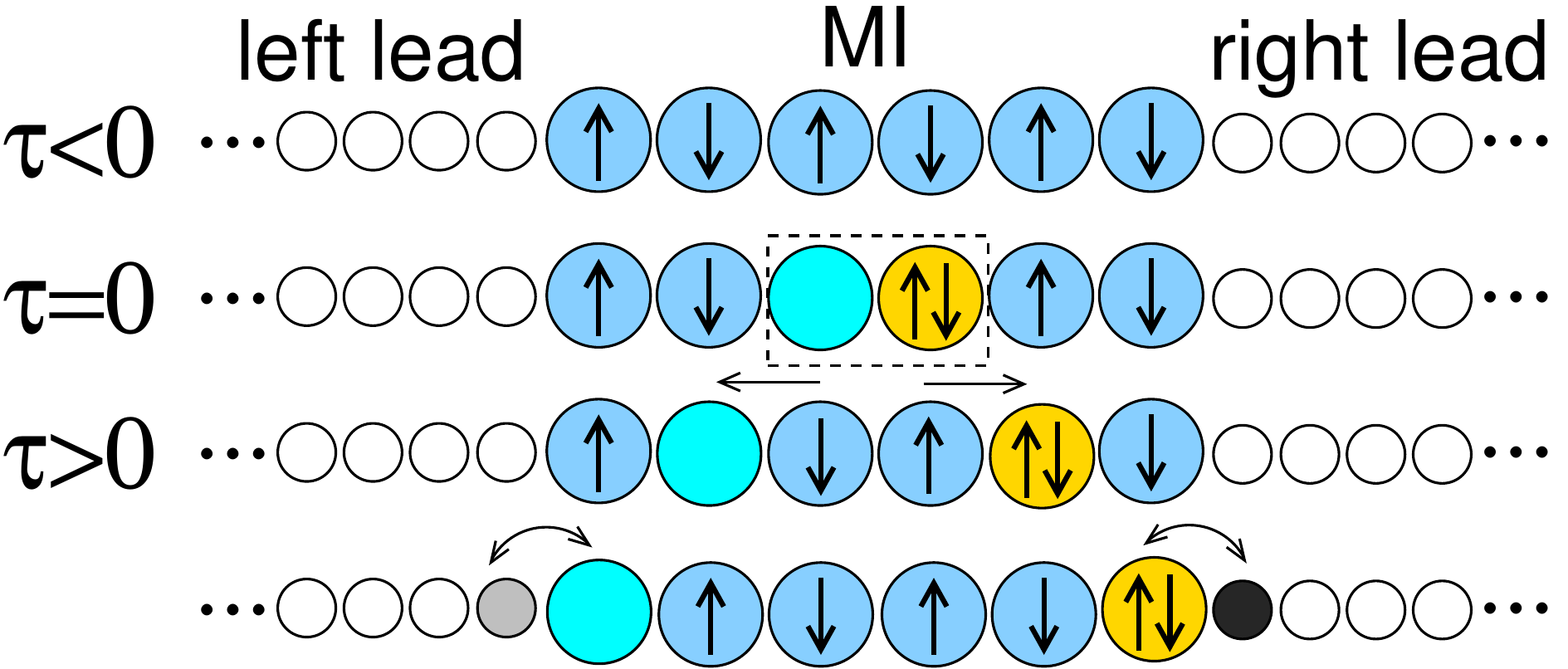}
\vspace{-0.5cm} \caption{ Schematic representation of the MI+leads system studied here.
At time $\tau=0$, a
doublon-holon pair is created at the center of the MI region. This pair propagates and
eventually reaches the boundary interfaces.  } \label{fig:ChainExciton}
\end{figure}

The main results are summarized as follows. A strong confinement of the
holon-doublon pairs within the MI region occurs for moderate to high values of
the on-site interaction $U$. Nevertheless, charge transmission through the
MI-metal boundary can be favored by properly adjusting the voltage difference
between the metallic leads. These dynamical effects have a strong influence on
the spin-spin correlations within the MI region, effectively reducing the
power-law decaying antiferromagnetic (AFM) order and introducing phase shifts
in the spin structure factor.

The paper is organized as follows. The model and the details of the calculation
are described in Sec. \ref{sec:model-methods}. The main results are presented
in Sec. ~\ref{sec:results}, where we discuss the charge, double occupation, and
spin-spin correlation dynamics (Sec. \ref{sec:MIleadsChargetransfer}), as well
as compare results against noninteracting cases (Sec.
\ref{sec:comp_noninteracting}). The phase shifts in the spin structure factor
are also discussed (Sec. \ref{sec:shiftsSq}). A summary of the main results is
given in Sec.~\ref{sec:summary}.

\section{Model and Methods}
\label{sec:model-methods}

The metal-MI-metal junction studied here is described
by a 1D Hubbard model with on-site and
nearest-neighbor  Coulomb interaction terms (representing the MI region) connected to
noninteracting sites (representing the metallic  leads). The Hamiltonian reads
$H=H_{\MottIns}+H_{\leads}+H_{\coupling}$, with:

\begin{eqnarray}
H_{\MottIns}&=&-t^{\prime \prime} \sum_{\sigma,i=N_L+1}^{N_L+N_{\MottIns}} \!\!  c^{\dagger}_{i \sigma} c_{i+1 \sigma} + U\left(n_{i \uparrow}\!\!-\!\!\frac{1}{2}\right)\left(n_{i \downarrow}-\frac{1}{2}\right)  \nonumber \\
 && + V \left(n_{i}-1\right)\left(n_{i+1}-1\right) + \mbox{h.c.} \nonumber \\
H_{\leads}&=& \sum_{i \in R,L} \mu_{R,L} n_i - t \sum_{\sigma,i \in R,L}c^{\dagger}_{i \sigma} c_{i+1 \sigma} + \mbox{h.c.} \nonumber \\
H_{\coupling}&=&-t^{\prime } \sum_{\sigma,i=N_L;i=N_L+N_{\MottIns}}\left( c^{\dagger}_{i \sigma} c_{i+1 \sigma} + \mbox{h.c.} \right)
\label{Eq:Hamiltonian}
\end{eqnarray}
where the sum in $H_{\coupling}$ has only two terms ($i=N_{\rm L}$ and
$i=N_{\rm L}+N_{\MottIns}$). As usual, $c^{\dagger}_{i \sigma}$ ($c_{i
\sigma}$) represents the creation (destruction) operator of an electron with
spin projection $\sigma$ at site $i$.

In the model above, $H_{\MottIns}$ describes the central interacting region, with $U$
and $V$ being, respectively, the on-site and nearest-neighbor Coulomb repulsion.
$t^{\prime\prime}$ is the hopping matrix element between sites in the MI region. The second
term ($H_{\leads}$) represents the noninteracting leads, where $t$ is the
tight-binding hopping amplitude (taken as the unit of energy hereafter) and $\mu_{\ell}$ is
the on-site diagonal energy at the sites forming the
lead $\ell$. Finally, $H_{\coupling}$ describes the coupling between the leads
and the MI involving a hopping amplitude $t^{\prime}$.

The total length of the 1D chain representing the entire system is $N_{\rm
L}+N_{\MottIns}+N_{\rm R}$, where $N_{\rm L}$, $N_{\MottIns}$, and $N_{\rm R}$
correspond to the number of sites in the left lead, the Mott insulator,
and the right
lead, respectively (referred to as a ``$N_{\rm L}$-$N_{\MottIns}$-$N_{\rm R}$
configuration''). The key parameters governing the  dynamics are the on-site
interaction  $U$ within the MI and the
on-site energy difference between the leads, given by $\Delta  \mu \equiv
(\mu_{\rm L}-\mu_{\rm R})/2$,
which can be controlled by applying different voltages to the leads
(for isolated leads in equilibrium at a fixed density, $\Delta \mu$ will be the
chemical potential difference between the leads).
In the following, we consider
symmetric chains with $N_{\rm L}=N_{\rm R}=20$ and $N_{\MottIns}=10$ (for a
total of $50$ sites for the system) and assume a strong lead-MI tunneling
amplitude ($t^{\prime}=t$), which minimizes reflections at the interface due to
a hopping matrix mismatch (although reflections due to the fact that $t^{\prime
\prime} \neq t$ still play a role).

To investigate the dynamics of the excitations and tunneling into the
metallic leads, we will focus on the regime of weakly bound doublon-holon pairs
with negligible recombination probability. Following Ref.
\onlinecite{Al-Hassanieh:166403:2008}, these constraints can be met by choosing
$U$ and $V$ such that $U/t^{\prime \prime} \gtrsim 8$ and $V/t^{\prime \prime}
\lesssim 0.6$. Thus, here we select the values $t^{\prime \prime}=0.5t$,
$V=0.3t$, and $U>4$ for our studies. As expressed before, all parameters with
units of energy hereafter are given in units of the hopping $t$.

Figure\ \ref{fig:ChainExciton}  depicts the metal-MI-metal junction. The
equilibrium (``$\tau<0$") ground-state $|\psi\rangle_0$ is obtained from static
density-matrix renormalization group (DMRG) (Refs.
\onlinecite{Schollwock:259:2005,Hallberg:477:2006}) calculations. At time
$\tau=0$, a doublon-holon pair  is suddenly created, by acting with holon
($h^{\dagger}_i$) and doublon ($d^{\dagger}_i$) creation operators  on the
state $|\psi\rangle_0$:
\begin{equation}
|\psi(\tau=0)\rangle=h^{\dagger}_p d^{\dagger}_{p+1}|\psi\rangle_0
\label{Eq:ExcitonCreation}
\end{equation}
with $h^{\dagger}_i\equiv  (1/\sqrt{2}) \sum_{\sigma} c_{i \sigma} (1-n_{i
\bar{\sigma}})$ and $d^{\dagger}_{i} \equiv (1/\sqrt{2})
\sum_{\sigma}c^{\dagger}_{i \sigma}n_{i \bar{\sigma}}$.

The site where the holon is created is   chosen   as $p=N_{\rm
L}+N_{\MottIns}/2$ so that the doublon-holon pair is initially localized at the
center of the MI region. This choice, although not  crucial, keeps the system
symmetric under the application of a particle-hole transformation followed by a
reflection through the middle bond.

The real time dynamics is obtained by time-evolving the original state using
the time-dependent DMRG method\cite{PhysRevLett.93.076401,DaleyJSM2004} and
obtaining $|\psi(\tau)\rangle=e^{-i \hat{H} \tau} |\psi(0)\rangle$. We find
that a Suzuki-Trotter decomposition with time steps $\delta \tau=0.05-0.1$ and
keeping $200$ states during the time evolution provides an adequate choice for
time-evolved quantities up to times $\tau \sim 40$. Since the sudden creation
of an exciton via Eq.\ \ref{Eq:ExcitonCreation} is restricted to only two sites
in the original system, we find that, starting from a well converged initial
state, increasing the number of states during the time evolution does not
significantly alter the main results.

To probe the effects of the doublon-holon dynamics on the
properties of the system, we calculate the expectation values of local operators
$\hat{{\mathcal O}}_i$ at site $i$ at each time step. We define the change from
the corresponding equilibrium value as $\delta \hat{{\mathcal O}}_i(\tau)\equiv
\langle\hat{{\mathcal O}}_i\rangle(\tau)-\langle\hat{{\mathcal O}}_i
\rangle_{\Equilib}$, where $\langle ... \rangle$ indicates the expectation
value using the time-evolved state $|\psi(\tau)\rangle$ at time $\tau$ while
$\langle\hat{{\mathcal O}}_i \rangle_{\Equilib}$ is the value calculated at
equilibrium (``$\tau<0$"). Typical site operators considered here are the
on-site charge $n_i$ and double occupation $D_i \equiv d^{\dagger}_i d_i$.

We also define the ``charge transfer'' to the right metallic lead as
\begin{equation}
\Delta n_{\rm R}(\tau, \Delta \mu) \equiv \sum_{i \in R} \langle n_i \rangle (\tau) - \langle n_i
\rangle_{\Equilib} = \sum_{i \in R} \delta n_i(\tau),
\label{Eq:ChargeSpinTrf}
\end{equation}
where the sum runs over sites in the right lead only. This quantity keeps track
of the time-integrated charge that is transported into the right lead as a result of
the creation of the holon-doublon excitation at $\tau=0$. A small charge transfer would indicate
strong confinement of the doublon-holon pair within the MI region.

We should point out that, as defined in Eq.\ \ref{Eq:ExcitonCreation}, the
doublon is created to the ``right" of the holon (i.e., a ``left/right"
holon-doublon pair), making the right lead the ``doublon side" and thus
justifying the choice of the right lead for the definition of the charge
transfer in Eq.\ \ref{Eq:ChargeSpinTrf}. This is only an arbitrary convention
since the creation of ``left/right" and ``right/left" holon-doublon pairs by
optical absorption should occur with equal probability. In fact, the effective
total charge transfer to the right metallic lead should take into account the
contribution from a ``right/left" holon-doublon pair as well. Given the
symmetries of the system described above, this contribution will be given by
$-\Delta n_{\rm R}(\tau, -\Delta \mu)$ calculated for the left/right holon-doublon
pair. Note that, once this contribution is taken into account, the net current
is nonzero only for $\Delta \mu \neq0$

\begin{figure}[!t]
\includegraphics*[height=0.7\columnwidth,width=1.0\columnwidth, bb=0 0 2500 1800,clip]{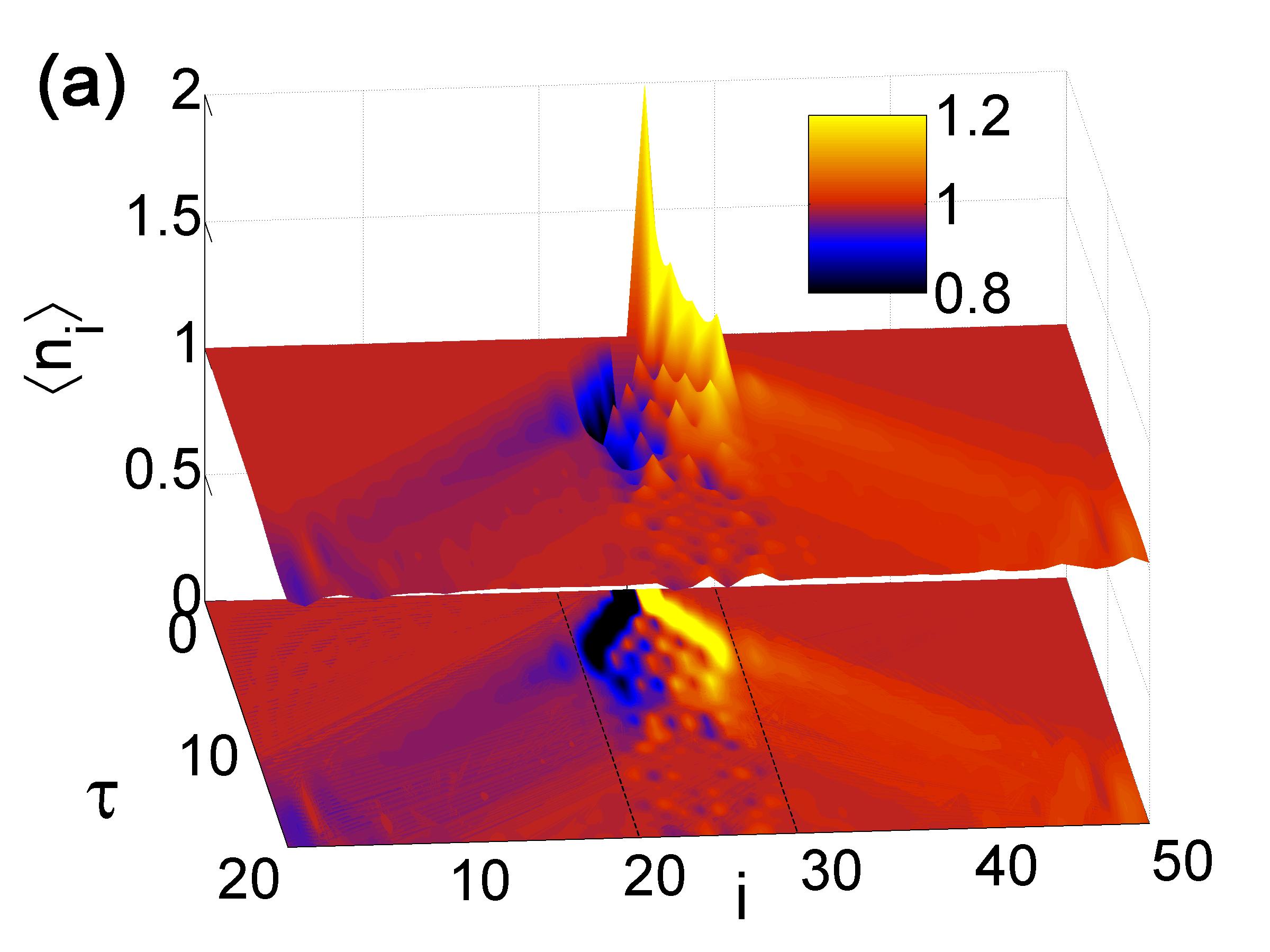}
\includegraphics*[height=0.7\columnwidth,width=1.0\columnwidth, bb=0 0 2500 1800,clip]{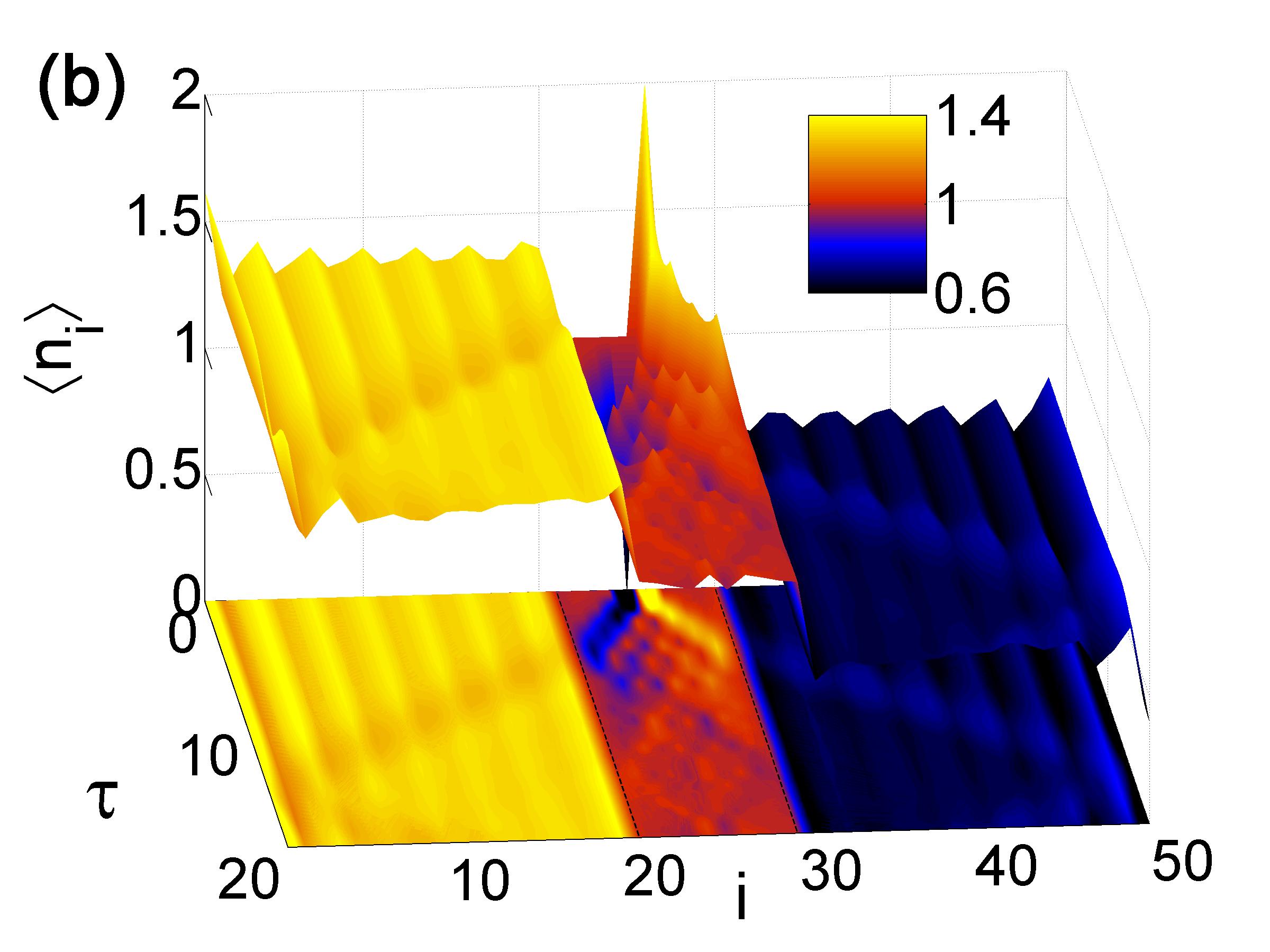}
\vspace{-0.5cm} \caption{  Charge $\langle n_i \rangle$ at site $i$ versus time
$\tau$ for a chain with $N=50$ sites (20-10-20 configuration), using $U=4$ and (a) $\Delta
\mu=0$ and
 (b) $\Delta \mu=-1$.  }
\label{fig:ChargeSiteVsTime}
\end{figure}

Additionally, we investigate the spin-spin correlations away from equilibrium.
We calculate the dynamical spin structure factor $S(q,\tau)$, defined as:
\beq S(q,\tau) = \frac{1}{N}\sum_{j,k \in \MottIns} e^{i(j-k)q} \langle
S^{z}_{j}S^{z}_{k}\rangle(\tau), \eeq
with $j,k$ spanning the interacting region ($j,k \in \MottIns$). As usual, a
peak in $S(q)$ at $q=\pi$ signals quasi-long-range antiferromagnetic order,
given in real space by $\langle S^{i}_{j}S^{z}_{j+k}\rangle  \sim
(-1)^{k}/r^{\alpha}$ with $\alpha$ being the  critical exponent on the order
parameter.

\section{Results}
\label{sec:results}

\subsection{MI-leads charge transfer}
\label{sec:MIleadsChargetransfer}

The typical dynamical behavior of the local charge $\langle n_i \rangle (\tau)$
is depicted in Fig.\ \ref{fig:ChargeSiteVsTime}-a for $U=4$ and $\Delta \mu=0$.
At $\tau<0$, the ground state of the system is particle-hole symmetric, with
$\langle n_i \rangle=1$ for all $i$. At $\tau=0$, the doublon-holon pair is
created, making $\langle n_p \rangle=0$, $\langle n_{p+1} \rangle=2$, and
$\langle D_p \rangle=0$, $\langle D_{p+1} \rangle=1$ (note that, while the
total charge $\sum_{i} n_i$ is conserved, $\sum_{i} D_{i}$ is not constant).
Due to the on-site repulsion in the MI, these charge excitations move in
opposite directions as a function of $\tau$, eventually reaching the MI-metal
interfaces at a time scale $\tau \sim \tau_r$ (for the parameters in Fig.\
\ref{fig:ChargeSiteVsTime}, $\tau_r \approx 5.5$).

\begin{figure}[!t]
\includegraphics*[height=1.0\columnwidth,width=1.0\columnwidth, bb=50 50 350 300,clip]{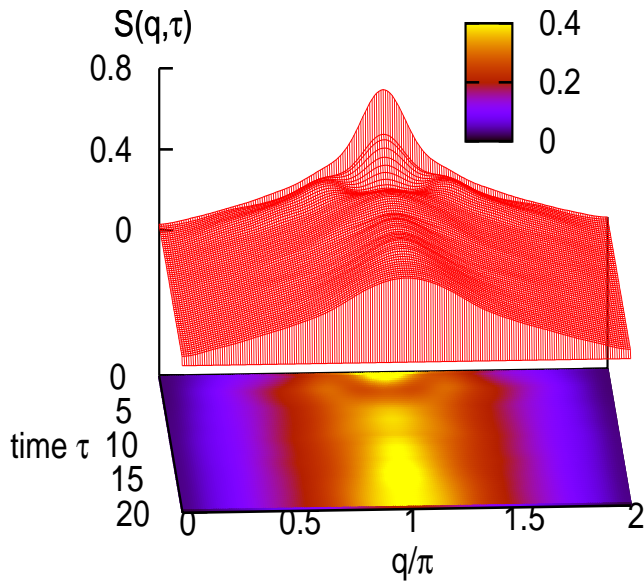}
%
%
%
%
\caption{  Spin structure factor $S(q,\tau)$ in the MI region for $U=4$ and $\Delta \mu=0$.}
\label{fig:Sq_U4}
\end{figure}

For $\Delta \mu \neq 0$, the
charge is expected to be non-uniformly distributed, with clear charge ``plateaus''  in each of
the three  regions. This is clearly depicted in Fig.\ \ref{fig:ChargeSiteVsTime}-b,
with $\mu_{\rm R}>\mu_{\rm L}$ ($\Delta \mu<0$). In this case, the equilibrium charge
distribution of the system is such that $\langle n_{i \in L} \rangle > 1$, and
$\langle n_{i \in R} \rangle < 1$ while $\langle n_{i \in \MottIns} \rangle
\sim  1$ in the MI region. As the doublon excitation approaches  the MI-leads
boundary, it  is then partially transmitted to the right lead while, by
symmetry, the holon excitation is partially transmitted to the left lead.

An intuitive picture for such a charge transmission/reflection can be
formulated in terms of the single-particle density of states. The energy cost
for the formation of the doublon is on the order of the Mott gap,
$\Delta_{\MottIns}$. Therefore, from purely energetic considerations, one
expects an enhanced charge transport if there are enough states available in
the leads at energies on the order of $\pm \Delta_{\MottIns}/2$ relatively to
the Fermi energy. Static DMRG calculations for the equilibrium density of
states (not shown) give Mott gap values $\Delta_{\MottIns}\approx 2$ for $U=4$
and $\Delta_{\MottIns}\approx 8$ for $U=10$, which is on the order or larger
than the typical half bandwidth $\sim 2t$ of the noninteracting chain.
Therefore, for a fixed $\Delta\mu$, one expects a decreasing charge transfer
with increasing $U$ as there are fewer states in the leads available at
energies $\sim \pm \Delta_{\MottIns}$. On the other hand, a nonzero $\Delta
\mu$  can increase the charge transfer by allowing a ``matching" of states
available for $|\Delta\mu| \sim \Delta_{\MottIns}/2$.

In fact, the reflection at the boundary can be reduced by changing
$\Delta \mu$,
as shown in Fig.\ \ref{fig:ChargeSiteVsTime}-b. In this case, the
Fermi energy in the leads is aligned with the band, creating states available
for the excitation to ``leak" into the leads.

\begin{figure}[!t]
\includegraphics*[height=0.99\columnwidth,width=1.0\columnwidth, bb=40 0 2450 1800,clip]{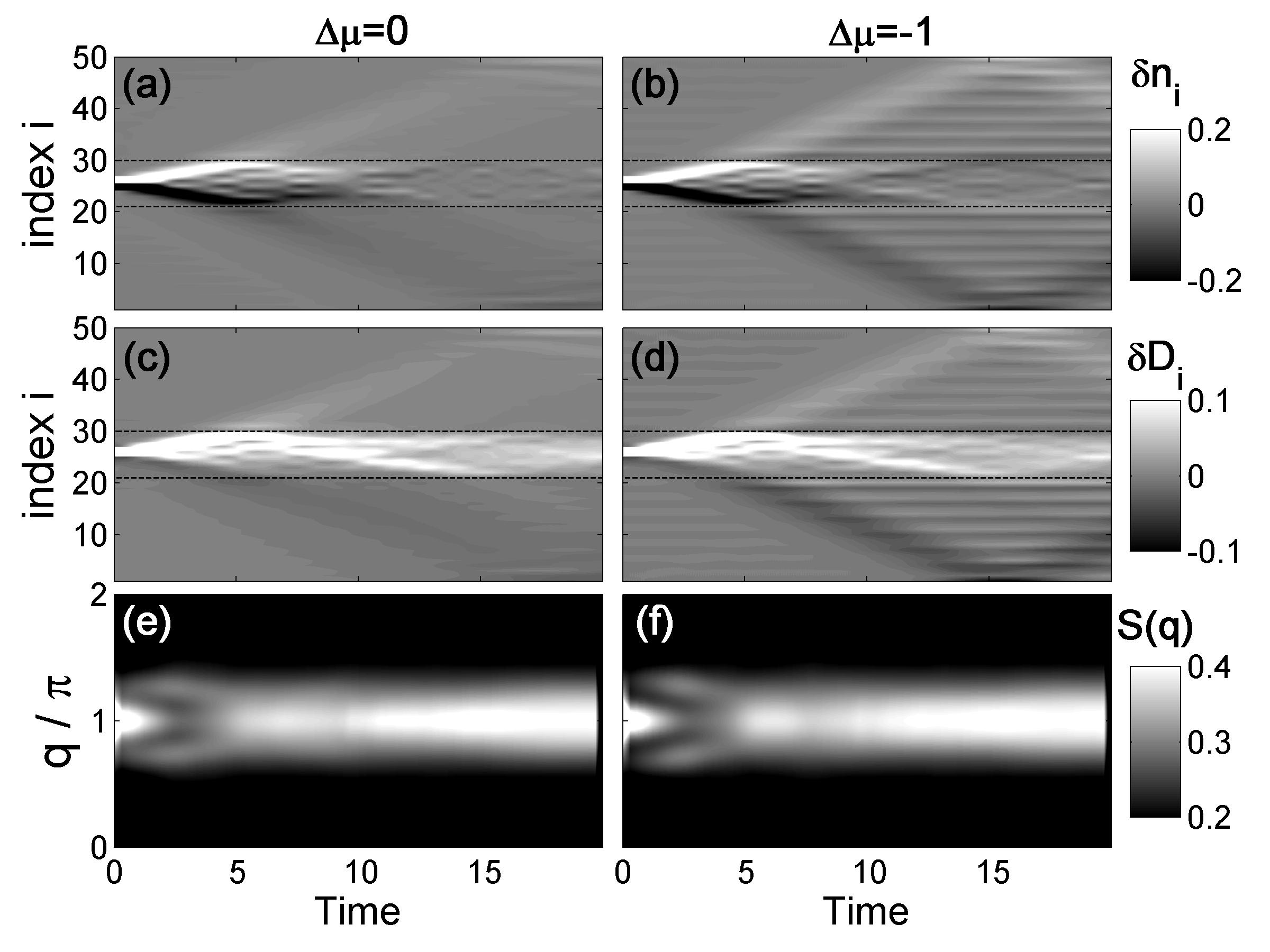}
\vspace{-0.5cm} \caption{  Nonequilibrium charge difference $\delta n_i$ (a,b),
double occupation $\delta D_i$ on each site (c,d), and spin structure
factor $S(q,\tau)$ in the MI region (e,f) versus time. The results are for $U=4$ with different
values of $\Delta \mu$:  $\Delta \mu=0$
(a,c,e) and
$\Delta \mu=-1$ (b,d,f).
Note the pronounced charge transfer for
$\Delta \mu=-1$.} \label{fig:ChargeVsTime_varyDmu}
\end{figure}

This simplified picture, however, does not take into account the existing
correlation effects within the Mott insulator region. Additional insight on the
effect of the doublon-holon dynamics can be provided by the spin-spin
correlations within the MI region, via the spin structure factor. Fig.\
\ref{fig:Sq_U4} shows the time evolution for $S(q,\tau)$ calculated within the
interacting (MI) region for $U=4$. In equilibrium (``$\tau<0$", shown as the
first curve), a pronounced  peak is observed at $q=\pi$ in both cases,
signaling AFM correlations.

At intermediate times, this peak acquires ``shoulders'' which become more
prominent around $\tau \sim \tau_r/2$, which coincides with the time scale for
which the doublon-holon pair is spatially separated but has not yet reached the
boundary. For longer times, the peak is broadened, indicating a weaker AFM
order within the MI region.
The shoulders in $S(q,\tau)$ are consistent with the formation of anti-phase
domain walls in the MI region due to the spatial separation of the doublon and
the holon, as we discuss further in Sec.\ \ref{sec:shiftsSq}. The fact that
these features in the spin structure factor closely track the charge dynamics
indicates that charge excitations (rather than collective spin excitations) are
the dominant decay channel for the doublon-holon pair, in accordance to
previous findings.\cite{Al-Hassanieh:166403:2008}
%

Figure \ref{fig:ChargeVsTime_varyDmu} presents a visual summary of the dynamics
of the charge, double occupation, and spin-spin correlations for $U=4$ and
different values of
$\Delta \mu$. Figs.\ \ref{fig:ChargeVsTime_varyDmu} a-b show $\delta n_i(\tau)$
at all sites for $\Delta\mu=0$ and $\Delta\mu=-1$, respectively. An enhanced
charge transmission into the leads is clearly seen for $\Delta\mu=-1$,
consistent with the arguments given above. This additional charge is an effect
of the doublon excitation tunneling into the leads. This is confirmed by
analyzing the change in double occupation from equilibrium $\delta D_i(\tau)$
(Fig.\ \ref{fig:ChargeVsTime_varyDmu} c-d). Note that the doublon is ``long
lived", as evidenced by the dynamics of the double occupation within the MI
region as it reflects off the boundary. This is a consequence of the weak decay
of the doublon for large $U/t^{\prime \prime}$,
as discussed in Ref.\ \onlinecite{Al-Hassanieh:166403:2008}.

Additionally, there are interesting effects in the spin-spin correlation, as
shown in the  $S(q,\tau)$ plots in Figs.\ \ref{fig:ChargeVsTime_varyDmu} e-f. A
clear splitting of the peak occurs as the doublon and holon excitations become separated
within the MI, also seen in Fig.\ \ref{fig:Sq_U4}. More interestingly is the
fact that this feature is enhanced for
$\Delta \mu < 0$
(Fig.\ \ref{fig:ChargeVsTime_varyDmu}-f),
a consequence of having a stronger
charge tunneling between the MI and the leads (i.e. stronger ``doping", as
discussed in Sec.\ \ref{sec:shiftsSq}).

\begin{figure}[!t]
\includegraphics*[height=0.7\columnwidth,width=1.0\columnwidth, bb=40 180 590 620,clip]{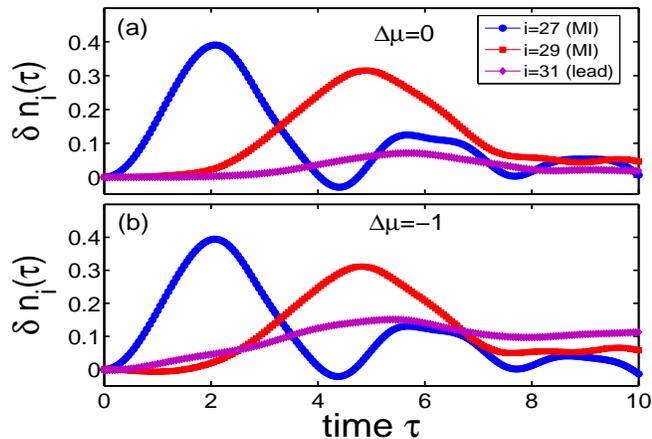}
\vspace{-0.5cm} \caption{  Charge difference $\delta n_i(\tau)$ on sites near
 the MI-metal boundary as a function of time for $U=4$ and (a) $\Delta \mu=0$
 and (b) $\Delta \mu=-1$. }
\label{fig:ChargeInterface}
\end{figure}

A closer look on the charge dynamics at the interface is presented in   Fig.\
\ref{fig:ChargeInterface},   which  depicts  the behavior of  the added  charge
$\delta  n_i(\tau)$ at sites on both sides of the interface. A weak odd-even
modulation is present so only odd-numbered sites are shown for a meaningful
comparison. The amplitude of the charge excitation decays as the ``charge
front" approaches the boundary. In addition, a reflection at the interface
takes place for $\tau \sim \tau_r=5.5$ in both cases.

We  can    estimate the   MI-metal charge reflection by comparing the maximum
amplitudes  of $\delta n_i(\tau)$   at sites on each side of the interface. For
$\Delta \mu\!\!=\!0$ (Fig.\ \ref{fig:ChargeInterface}-a), the maximum in
$\delta n_i$ drops   from $0.315$ for  $i\!\!=\!\!29$  (within the  MI) to
$0.071$ for $i\!\!=\!\!31$ (in the leads),  giving
$R\!\!\equiv\!\!\left(\mbox{max }\delta n_{29}-\mbox{max }\delta
n_{31}\right)/\mbox{max }\delta n_{29}\!\!=\!\!0.7737$.  The relative  drop is
smaller for $\Delta \mu\!\!=\!-1$ (Fig.\ \ref{fig:ChargeInterface}-b), with
$R\!\approx\!0.51$.

\begin{figure}[!t]
\includegraphics*[height=1.0\columnwidth,width=0.9\columnwidth,bb=4 1 787 610,clip]{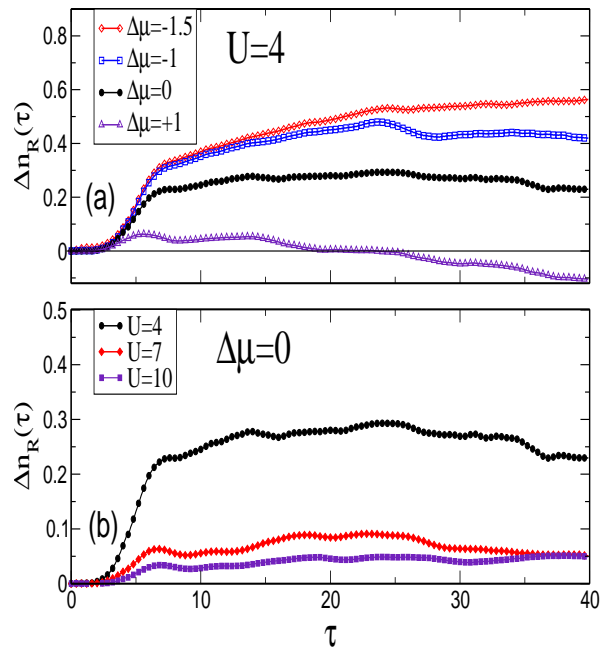}
%
%
\vspace{-0.2cm} \caption{ Charge transfer  (Eq.\ \ref{Eq:ChargeSpinTrf}) for
(a) $U=4$ and different values of $\Delta\mu$ and (b) $\Delta\mu=0$ and different values of $U$, with $t^{\prime\prime}=0.5$ in all cases.}
\label{fig:ChargeSpinTransfer_vsDmu}
\end{figure}

For longer times,   further scattering of  the doublon  and  holon excitations
off the MI-metal interface result in additional charge tunneling into the
leads. The integrated charge transferred to  the right lead  ($\Delta n_{\rm R}$,
defined in Eq.\ \ref{Eq:ChargeSpinTrf})   is plotted in   Fig.\
\ref{fig:ChargeSpinTransfer_vsDmu}-a for  different values  of $\Delta \mu$.
For longer   times,  $\Delta n_{\rm R}(\tau)$ reaches an  approximate plateau, indicating a
steady state of the charge in the lead (for $\Delta\mu=-1.5$ the plateau has
not been fully reached on the maximum time used in the calculations). As
expected from our previous results, the height of the plateau is larger for
negative values of $\Delta \mu$. For a stronger interaction $U$ (larger
$\Delta_{\MottIns}$), the charge transfer is strongly suppressed, as shown in
Fig.\ \ref{fig:ChargeSpinTransfer_vsDmu}-b.

\begin{figure}[!t]
\includegraphics*[height=0.99\columnwidth,width=1.0\columnwidth, bb=40 0 2450 1800,clip]{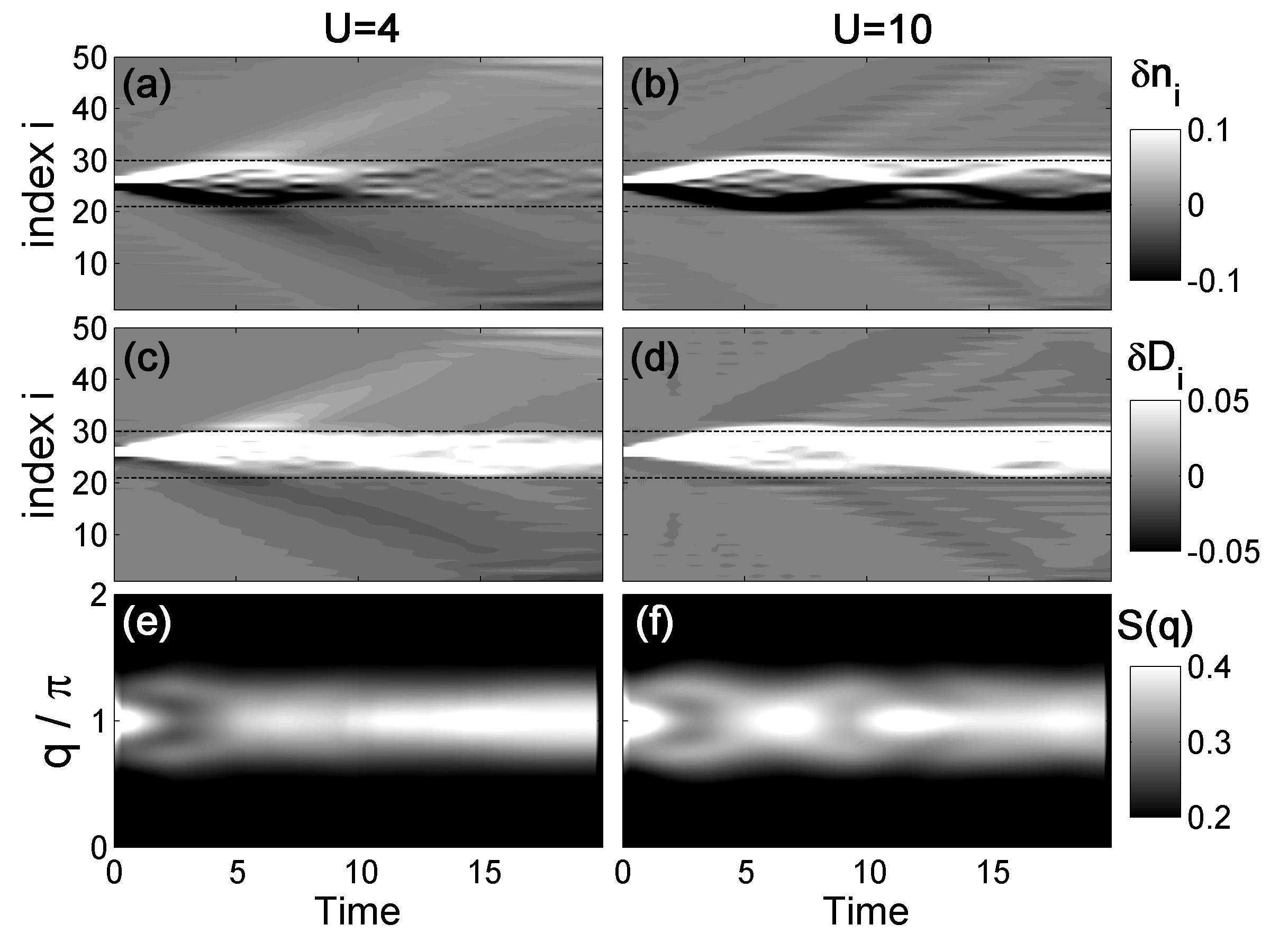}
\vspace{-0.5cm}
\caption{  Nonequilibrium charge $\delta n_i(\tau)$ (a,b) and double occupation
$\delta D_i(\tau)$ (c,d) on each site as a function of time for $U=4$ (a,c) and
$U=10$ (b,d). Panels (e,f) show the corresponding spin structure factor
$S(q,\tau)$ in the MI region.}
\label{fig:ChargeVsTime_varyU}
\end{figure}

These results indicate that the stronger the interaction $U$ in the MI region,
the more confined the holon-doublon pair becomes. This is clearly seen in
Figs.\ \ref{fig:ChargeVsTime_varyU}(a,b) and (c,d), which depicts contour plots
of $\delta n_{i}(\tau)$ and $\delta D_i(\tau)$ for $U=4$ and $U=10$,
respectively. For $U=10$, most of the charge remains confined in the MI region,
reflecting off the interface in a periodic pattern. Interestingly, the
long-lived holon and doublon excitations also ``repel" each other, as seen for
$\tau \sim 2\tau_r$.

Not surprisingly, increasing the Mott gap has a pronounced effect on the spin
structure factor $S(q,\tau)$. Figs.\ \ref{fig:ChargeVsTime_varyU}-e and
\ref{fig:ChargeVsTime_varyU}-f show $S(q,\tau)$ for $U=4$ and $U=10$,
respectively. For $U=4$, the three-peak structure (``shoulders") disappear for
$\tau \gtrsim \tau_r$, where a broad peak takes over. This is about the time
scale for which the charge/spin fronts reach the boundary (Fig.\
\ref{fig:ChargeVsTime_varyU}-b). For $U=10$, the shoulders reappear at later
times, consistent with the ``beating'' in the doublon/holon pair as they
reflect back and forth within the MI region and with each other. This is
coupled with a much stronger confinement of the holon-doublon pair within the
MI region, as seen in Figs.\ \ref{fig:ChargeVsTime_varyU}-b.

\subsection{Comparison with the non-interacting case}
\label{sec:comp_noninteracting}

An important question is how the results for the MI-leads charge transfer
compare to a case where no electron-electron interactions are present in the
central region. To address this question, we present results for two
distinct cases: (i) a full ``metallic" system, i.e., $U=V=0$ and
$t^{\prime\prime}=0.5$ in the $H_{\MottIns}$ term in Eq.\ \ref{Eq:Hamiltonian}
and (ii) a ``band insulator" region, in which $H_{\MottIns}$ is replaced
by
\begin{equation}
H_{\BandIns}=-t^{\prime\prime} \sum_{\sigma, i=N_L+1}^{N_L+N_{\MottIns}} \!\! \left(
1+\delta(-1)^{i}\right)c^{\dagger}_{i \sigma} c_{i+1 \sigma} + \mbox{h.c.}
\end{equation}
in the Hamiltonian. Notice that $H_{\BandIns}$ represents a Peierls chain,
which has a charge gap for $\delta \neq 0$ and the gap size will depend on
$t^{\prime\prime}$ and $\delta$. In the following, we choose
$t^{\prime\prime}=1$ and $\delta=0.6$ so that the band gap is nearly the same
as the Mott gap for the case shown in Fig.\ \ref{fig:ChargeSiteVsTime}-a
($U=4$, $V=0.3$, and $t^{\prime \prime}=0.5$).

\begin{figure}[!t]
\includegraphics*[height=0.7\columnwidth,width=1.0\columnwidth, bb=0 0 2500 1800,clip]{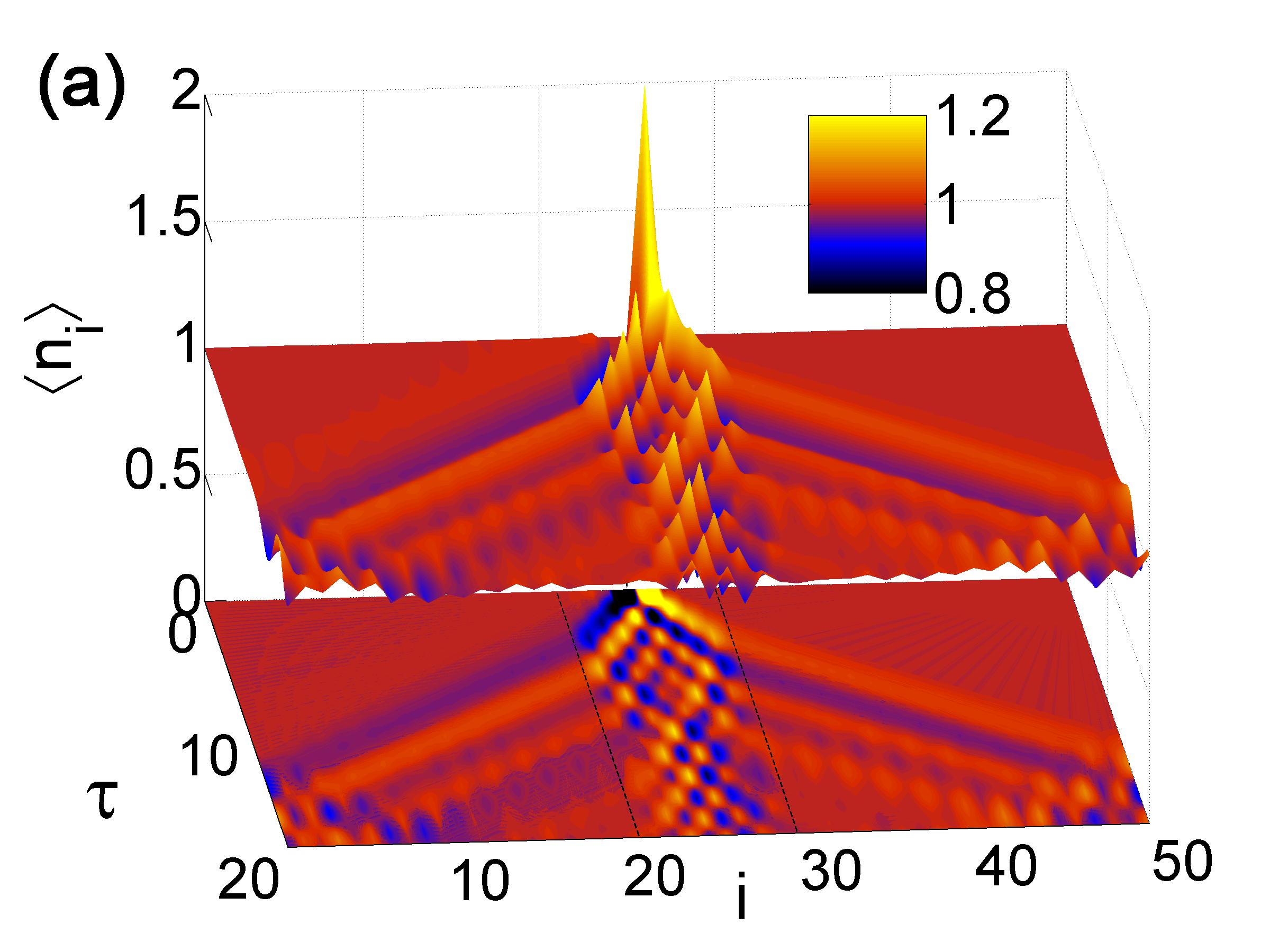}
\includegraphics*[height=0.7\columnwidth,width=1.0\columnwidth, bb=0 0 2500 1800,clip]{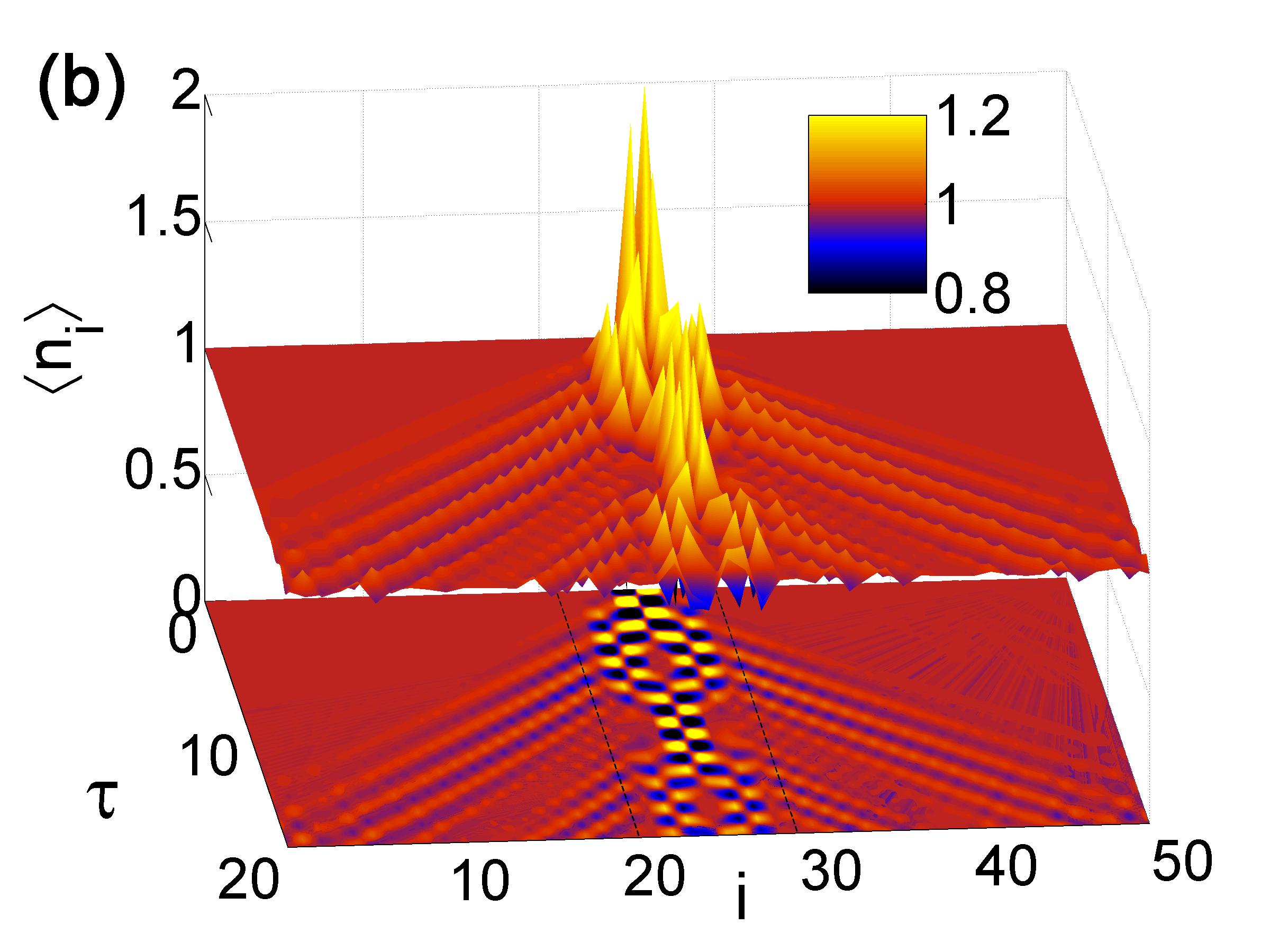}
\vspace{-0.5cm} \caption{  Charge $\langle n_i \rangle$ on site i versus time
$\tau$ for $N=50$ sites, a non-interacting central region ($U=0$), and $\Delta \mu=0$. (a)
is using a Hamiltonian as in Eq.\ \ref{Eq:Hamiltonian} with $t^{\prime \prime}=0.5$ while in (b)
$H_{\rm BI}$ replaces $H_{\MottIns}$ and $t^{\prime \prime}=1$, $\delta=0.6$ (band
insulator).} \label{fig:ChargeSiteVsTimeNoninteracting}
\end{figure}

Figures\ \ref{fig:ChargeSiteVsTimeNoninteracting}-a and b show the occupation
$\langle n_i \rangle$ on the chain sites for the fully metallic and band
insulator cases, respectively. The first case corresponds to the noninteracting
limit of the results presented in the previous section, with a fully metallic
chain. There are still strong reflections at the boundary due to the hopping
mismatch $(t^{\prime\prime} \neq t)$. More importantly, the holon-doublon
charge excitations produce Friedel-type oscillations along the chain, forming a
clear charge interference pattern over time (see contour plot in Fig.\
\ref{fig:ChargeSiteVsTimeNoninteracting}-a).

For the case where the central region is a band insulator, charge oscillations
also occur, leading to a ``checkerboard" pattern of alternating positive and
negative charges in the central sites (Fig.\
\ref{fig:ChargeSiteVsTimeNoninteracting}-b). In this case,  propagation of a
``charge wavefront" is clearly suppressed as compared to the fully metallic
case.

\begin{figure}[!t]
%
%
\includegraphics*[height=0.9\columnwidth,width=1.0\columnwidth,bb=4 3 787 605,clip]{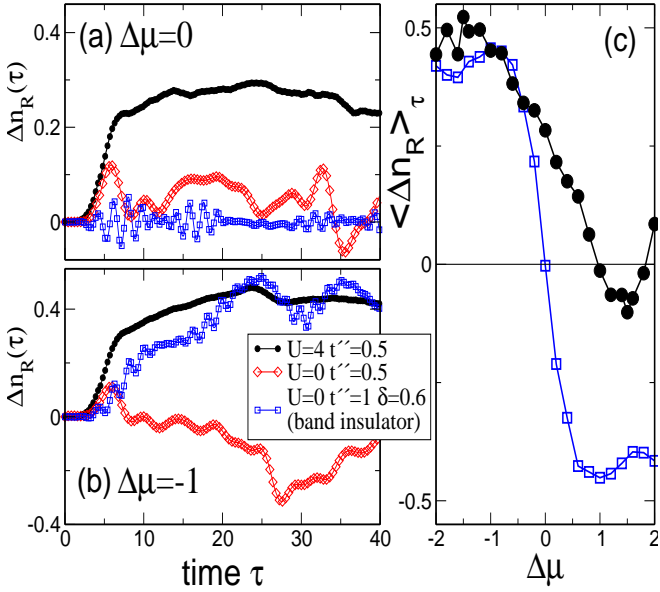}
%
\vspace{-0.5cm} \caption{ Charge transfer comparison between the cases of interacting ($U$=4)  and
noninteracting ($U$=0) central regions. Results are also shown for a band insulator (Peierls chain)
with a gap very similar to that of the Mott insulator with $U$=4.
(a,b) $\Delta n_R(\tau)$ versus $\tau$ for
$\Delta \mu=0$ and $\Delta \mu=-1$, respectively.
(c) Time-averaged (steady-state) charge transfer $\langle \Delta n_R \rangle_{\tau}$ versus $\Delta \mu$ for the MI and BI cases.
}
\label{fig:ChargeTrfCompareUeq0}
\end{figure}

The charge transfer to the right lead in both situations is clearly distinct
from the interacting case. Figs.\ \ref{fig:ChargeTrfCompareUeq0}-a,b show a
comparison of the charge transfer at $U=4$ (same as depicted in Fig.\
\ref{fig:ChargeSpinTransfer_vsDmu}) with results at $U=0$ for $\Delta \mu=0$ and
$\Delta \mu=-1$, respectively.

The charge transfer is very limited for $\Delta \mu=0$ in the noninteracting
cases, having essentially a zero time-average for the band insulator. By
contrast, the case $U=4$ shows a positive transfer even for $\Delta \mu=0$ due
to the on-site repulsion within the central region. Note that the charge gap in
the central region is nearly the same for both the Mott and band insulator
cases, while the reflection at the boundary is much more accentuated for the
latter. This highlights the role of the electron-electron interactions in the
additional charge transfer for the $U=4$ case.
For $\Delta \mu=-1$ (Fig.\ \ref{fig:ChargeTrfCompareUeq0}-b) the charge
transfer is improved in the noninteracting cases, as expected. Overall, the
charge transfer is still more effective in the case where the central region is
a Mott insulator, as compared to the noninteracting cases.

In fact, this holds for a wide range of values of $\Delta \mu$, as it can be
seen in Fig.\ \ref{fig:ChargeTrfCompareUeq0}-c, where a time-averaged charge
transfer $\langle \Delta n_R \rangle_{\tau}$ (calculated by taking averages of
$\Delta n_R(\tau)$ over a time interval $\Delta \tau \sim 10$ at longer times,
away from the initial transient) is plotted as a function of $\Delta \mu$. In
the MI case, $\langle \Delta n_R \rangle_{\tau}$ shows a broad peak, over a
wide range of $\Delta \mu$ (larger than the typical metallic half-bandwidth,
$\sim 2$ in the units used), showing a positive charge transfer to the right
lead even for \textit{positive} values of $\Delta \mu$ (i.e., $\mu_R<\mu_L$).
This is in sharp contrast with the band insulator case, for which $\langle
\Delta n_R \rangle_{\tau} < 0$ for $\mu_R < \mu_L$. Notice that, for $\Delta
\mu=0$, $\langle \Delta n_R \rangle_{\tau}>0$ in the MI case while $\langle
\Delta n_R \rangle_{\tau}\approx0$ in the BI case (Fig.\
\ref{fig:ChargeTrfCompareUeq0}-c). The latter is an expected result: in a
noninteracting system with an initially uniform charge distribution, an
electron-hole excitation will produce charge oscillations about the initial
charge value (see Fig.\ \ref{fig:ChargeSiteVsTimeNoninteracting}) but, on
average, no charge transfer takes place.

\subsection{Phase shifts in S(q)}
\label{sec:shiftsSq}

To understand the three-peak feature in the spin structure factor (e.g., Fig.\
\ref{fig:Sq_U4}), we have calculated $S(q)$ for a Hubbard chain with $N_{\rm
L}=N_{\rm R}=0$ and $N_{\MottIns}=40$ sites using static DMRG. We performed
several DMRG calculations targeting states with different number of electrons
$N$ and total spin projection $S_z$. More specifically, we considered
half-filling ($N=N_{\MottIns}$) and hole-doped ($N=N_{\MottIns}-N_{\rm h}$
where $N_{\rm h}$ is the number of holes) cases, with $S_z=0,1/2$, and
calculated the corresponding $S(q)$ for the different situations.

\begin{figure}[!t]
\includegraphics[height=0.9\columnwidth,width=0.9\columnwidth,bb=6 1 786 606,clip]{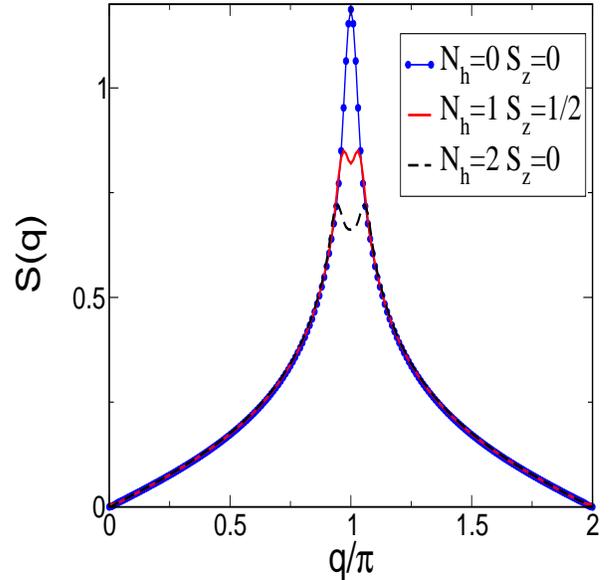}
%
\vspace{-0.5cm}\caption{ Equilibrium spin structure factor $S(q)$ for a
MI-only chain with $N$ sites (no leads) calculated for states with charge
$N-N_{\rm h}$ ($N_{\rm h}$ is the number of holes) and spin $S_z$. Parameters are $U=4$ and $t^{\prime \prime}=0.1$. }
\label{fig:HolesSpinFlips}
\end{figure}

Results are presented in Fig. \ref{fig:HolesSpinFlips}. For $N_{\rm
h}\!\!=\!\!0$ and $S_z\!\!=\!\!0$, the state corresponds to the
antiferromagnetically ordered ground-state of the system at half-filling, with
a peak at $q\!\!=\!\!\pi$. We then targeted states  with different values of
$N_{\rm h}$ and $S_z$. This corresponds to calculating the ground-state of a
system in which holes and/or spin  flips are added. For  $N_{\rm h}\!\!=\!\!1$,
$S_z\!\!=\!\!1/2$ (one hole) and $N_{\rm h}\!\!=\!\!2$,   $S_z\!\!=\!\!0$ (two
holes) a double peak structure appears.

This is reasonable  since holons will introduce anti-phase domain walls (ADW)
(i.e. ``$\pi$-shifts") in the AFM order and peaks at $\pi \pm \pi \nu_h$ are
expected (where $\nu_h$ is the hole density). As spin-flips are added (for
instance, $N_{\rm h}=1$, $S_z=3/2$) the double-peak becomes four peaks (not
shown). In light of these static calculations, the appearance of ``shoulders"
in $S(q,\tau)$  for $\tau<\tau_r$ (where $\tau_r$ is the time it takes for the
excitations to reach the boundaries) can be accounted for as follows.

The creation of the holon/doublon pairs at $\tau=0$ effectively removes two
magnetic moments from the MI region, making $S^z_{p (p+1)}=0$ at the holon
(doublon) site. This explains the decrease in the area under the $S(q,\tau)$
curve at $\tau=0$ as $\int S(q,\tau) dq \propto \sum_{i \in \MottIns} \langle
\left(S^z_i\right)^2 \rangle(\tau)$. For small $\tau$, $S(q,\tau)$ still
retains a single-peak structure initially since holon and doublon are on
adjacent sites and the corresponding phase shifts from ADWs cancel out.

For $\tau<\tau_r$, the MI region can be divided into two parts with magnetically
distinct characteristics: (i) an ``undoped" region where neither the doublon or holon excitations
have arrived and which still retains AFM order and (ii) a ``doped" region,
which has been already ``covered" by either the doublon or the holon
excitations. Part (i) contributes to a peak at $q=\pi$ in $S(q, \tau)$ while
part (ii) contributes to the two shoulders, as we expect from the static
calculations shown in Fig. \ref{fig:HolesSpinFlips} with $N_{\rm h}=2$.

For $\tau \sim \tau_r$, the excitation (and the corresponding ADW) reaches the
boundaries, making the phase shifts to cancel out again, suppressing the side
peaks. As the holon and doublon reflect off the boundaries, the side peaks
reappear up to time scales of order $\tau \sim 2\tau_r$. At this time scale,
doublon and holon excitations are again at nearest-neighbor sites and the
respective ADW phase-shifts cancel once again, leading to a single-peak
structure. The process then repeats up to time scales on the order of the
doublon decay time. Thus, this cycle is observed more clearly for larger values
of $U$ (e.g., Fig.\ \ref{fig:ChargeVsTime_varyU}-f) for which the doublon decay
time is large enough. Notice that $S(q,\tau \sim 2\tau_r)$, although featuring
a single peak at $q=\pi$,  has a much smaller area than $S(q,\tau=0)$,
indicating that the propagation of the holon-doublon pair significantly
modifies the AFM correlations.

\section{Summary}
\label{sec:summary}

In summary, we have studied the real-time propagation of doublon-holon
excitations in a Mott insulator (MI) connected to metallic leads. We analyze
the dynamics of charge, double occupation, and spin-spin correlations within
the MI, as well as the MI-leads charge transfer.

Our results indicate that the sharp change in the Hamiltonian at the MI-metal
interface hinders the charge transfer, suggesting that metals that closely
resemble the structure of the MI (Ref. \onlinecite{Kishida:929:2000}) would be
required as charge collectors instead of standard doped semiconductors. More
specifically, we find that the charge transfer across the MI-metal boundary is
quite sensitive to microscopic parameters in the MI region, particularly the
on-site interaction $U$. While $U$ needs to be sufficiently large so that the
doublon decay-time is larger than the typical time scale it takes to reach the
boundary, doublon-holon tunneling into the leads is suppressed for very large
values of $U$.

We believe two factors contribute to these results: (i) the increase in the
Mott gap and (ii) the fact that, for large $U$, the nature of the doublon and
holon excitations within the MI becomes more different than standard electron
and hole excitations in the noninteracting case. We have tested these
hypothesis by comparing configurations with either a Mott insulator or a
(noninteracting) band insulator in the central region: they show clear
differences in the charge transfer.

The repulsive interaction within the Mott insulator region favors the transfer
of the excess charge into the metallic leads even at $\Delta \mu=0$. This is in
sharp contrast with a noninteracting band insulator connected to leads with the
same band-gap for which the net charge transfer is essentially zero at zero
voltage difference.

Noninteracting and interacting cases show clearly distinct charge dynamics
after the excitation. In the former case, Friedel-type charge oscillations in
the central region are prominent at small times, while they are suppressed in
the interacting case and a clear spatial separation between hole-like and
particle-like excitations occurs.

Moreover, the propagation of holons and doublons within the MI region
dynamically alters the  AFM spin-spin correlations. In particular, extra
``$\pi$ shifts" appear in the spin correlation functions as doublons and holons
are spatially separated. We believe these qualitative findings will help on the
prospect of making future solar cell devices using strongly correlated
materials.

An interesting aspect that remains open is the effect of a finite temperature
in our results, a difficult problem considering that the numerical study of
transport in correlated systems at finite temperatures remains a very
challenging subject. The energy scales of spin and charge excitations in 1D
Mott insulators can be quite different ($\sim t^{\prime \prime}$ for gapped
charge excitations and $\sim (t^{\prime \prime})^2/U$ for gapless spin
excitations) and finite temperatures can affect each of these channels
differently. \cite{Fiete:801:2007,Halperin:81601:2007} In our case, however,
since the doublon-holon pairs decay very weakly into spinon modes and the
dynamics is governed by charge excitations, we expect the results to hold as
long as temperatures are small compared to the charge gap.

\acknowledgments

We acknowledge motivating discussions with Ivan Gonzalez, Fabian
Heidrich-Meisner, Hiro Onishi, and Satoshi Okamoto.
Research sponsored by the Division of Materials Sciences and
Engineering, Office of Basic Energy Sciences, U.S. Department of Energy.
Computational
support was provided by the National Energy Research Scientific Computing
Center (NERSC). Work at LANL was carried out under the auspices of the NNSA of
the U.S. DOE under Contract No. DE-AC52-06NA25396.
LGGVDS and ED acknowledge
support from the National Science Foundation via grant DMR-0706020.

\end{document}